# Practical biological spread-out Bragg peak design of carbon beam

Chang Hyeuk Kim, Hwa-Ryun Lee, Seduk Chang, Hong Suk Jang, Jeong Hwan Kim, Dong Wook Park, Won Taek Hwang, Tea-Keun Yang*

Korea Institute of Radiological and Medical Science,
Nowon-gil 75, Nowon-gu, Seoul 139 – 706

## Abstract

The carbon beams show more advantages on the biological properties compared with proton beams in radiation therapy. The carbon beam shows high linear energy transfer (LET) to medium and it increases the relative biological effectiveness (RBE). To design spread-out Bragg peak (SOBP) of biological dose using carbon beam, a practical method was purposed by using the linear-quadratic (LQ) model and Geant4 based Monte Carlo simulation code. The various Bragg peak profiles and LET was calculated for each slice at the target region. To generate appropriate biological SOBP, a set of weighting factor, which is a power function in terms of energy step, was applied to the obtained each physical dose. The designed biological SOBP showed 1.34 % of uniformity.

# I. Introduction

Using protons and heavier ion for medical treatment was purposed by Robert Wilson in 1946[1]. And, the first patient was treated with proton at the University of California at Berkeley in 1954[2]. The patient treatments using carbon ion have been performed mainly at GSI (German) and HIMAC (Japan). The current statistics shows 122449 patients have been treated by particle therapy in worldwide, 86.3 % (105,743) of the patients were treated by proton and 10.7 % (13,119) by carbon [3].

The energetic ion beams are generating Bragg peaks while they transfer the energy on the medium [4]. The characteristics of ion beam give a good advantage on radiation therapy. It gives low radiation dose at entrance region while give the maximum dose at target region. Compared with other light ions, the carbon is classified high linear energy transfer (LET) radiation. The high LET radiation led to higher relative biological effectiveness (RBE) then other, such as photon and proton [5]. In addition, the carbon generated Bragg peak has a steep distal fall-off which can be provided conformal dose delivery to avoid unwanted dose at critical organs.

Two different groups mainly performed the research regarding on the calculation of carbon RBE. One group develop Linear Quadratic (LQ) Model, which is based on the experiment results of reference cell line [6]. And, the other group suggested the Local Element Model (LEM), which simulated and calculated the effects of radiation from the spatial distribution of DNA double stand break [7], [8]. But, the RBE is varying by LET, biological end point and type of tissue or cell, etc. Therefore, it could have some difficulties to represent all possible options at the simulation process yet. Using the experiment result could be considered more practical way to calculate RBE value.

To design the spread-out Bragg peak (SOBP) of carbon, the experimented target cell response, which is survival curve depending on the different LET, is required. Because of the absence of actual carbon beam in Korea, the HIMAC experiment data, which is using human salivary grand (HSG) cell results was adopted [9], and also LQ model was used for RBE calculation at this study.

## II. Method

II.1 Dose and LET calculation

To design a Spread-out Bragg peak, the depth-dose profiles of various carbon beam energy should be calculated as well as LET. At this study, the Geant4 hadron therapy example was adopted [10]. The example code could be simulated passive beam line and active scanning beam line also. The code was known to model the eye therapy line on the INFN. In case of this study, only the response on the water phantom was focused. The physics model in this hadrontherapy example was recommended using QGSP_BIC_EMY as reference physics list. Quark Gluon String Precompound(QGSP) defines the hadronic models for nucleons. Binary Ion Casecade(BIC) defines the inelastic models for ions and ElectroMagnetic Y(EMY) defines the electromagnetic models for all particles [11]. The beam energies were selected 340, 370, 400, 430 MeV/u. The water phantom was defined 40x40 cm2 of transverse area and 40 cm of longitudinal length. The output data of simulations was collected each 0.1 mm of spacing in longitudinal direction of the phantom. The code generates two outputs; depth dose profile and linear energy transfer (LET) of carbon. The incident beam was defined as a pencil beam, which has 3 mm of radius and sigma is 2 mm. The beam energy was set to 0.1 % of energy variation, dE/E. The output of simulation was shown in figure 1.

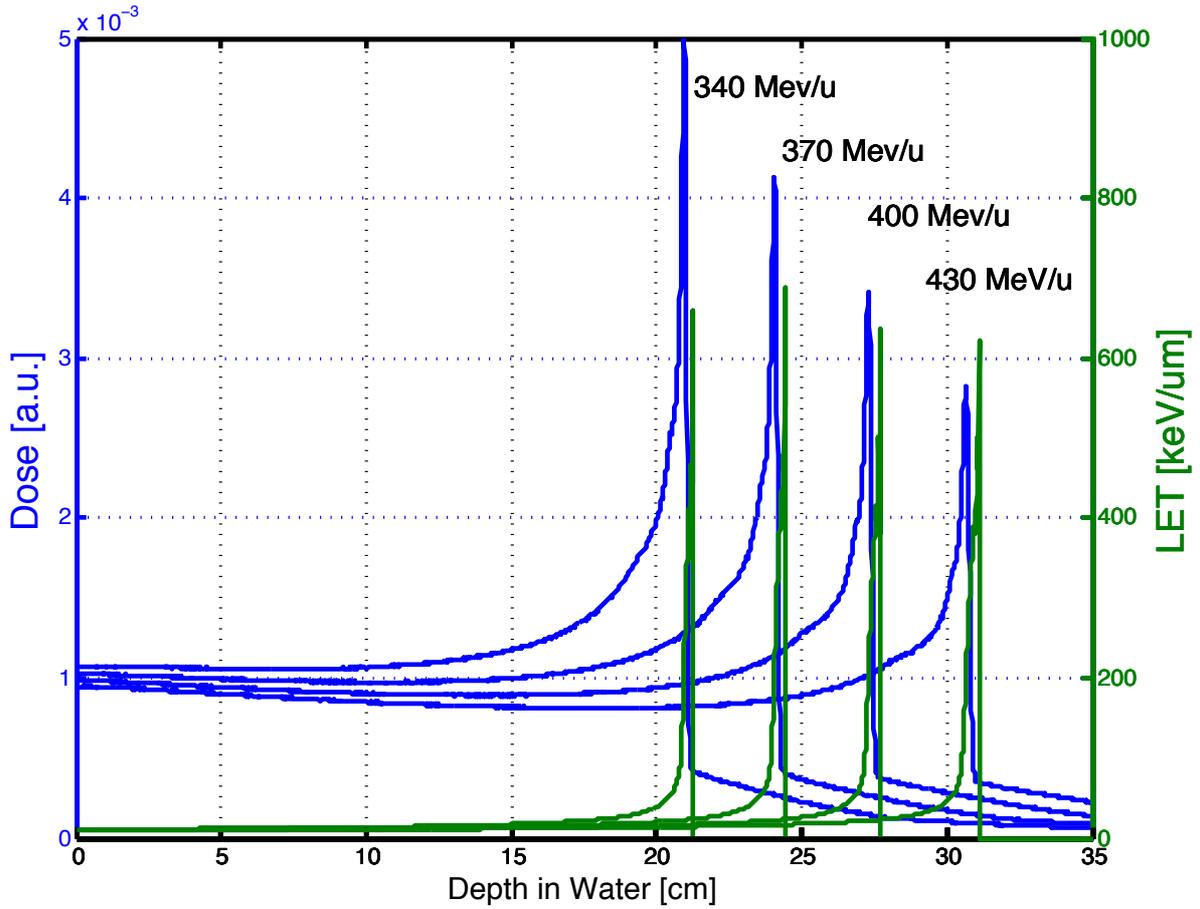

Fig 1. Obtained the depth dose profile and LET of various carbon energies

II.2 LQ model and RBE calculation

The cell killing effects are induced by several physical and bio-chemical processes. The LQ model is a mechanistic model of cell killing effects [12],which is related with repairing DNA process by double stand break and binary mis-repair of DSB from different radiation track. Simply, the effects based on the direct radiation will be shown in proportion to dose linearly, which is indicated as alpha parameter. But, the effect from indirect radiation are shown as proportional to the square of dose, which is called beta parameter. Therefore, it would be written as LQ formulation for the yield (Y) of legal lesion would be expressed by equation (1),

$$Y \propto \alpha D + \beta D^2 \qquad (1)$$

And, the assuming the Poisson distribution of lethal lesion, the survival fraction S would be expressed by equation (2),

$$S = \exp(-Y) = \exp[-(\alpha D + \beta D^2)] \qquad (2)$$

Therefore, the key parameter of LQ model is defining   and   as function of LET. At this study, the LQ model parameter was referred from NIRS [13]. The data was obtained based on the experimental result of human salivary gland (HSG) case. The fitted LQ parameters were shown in fig 2.

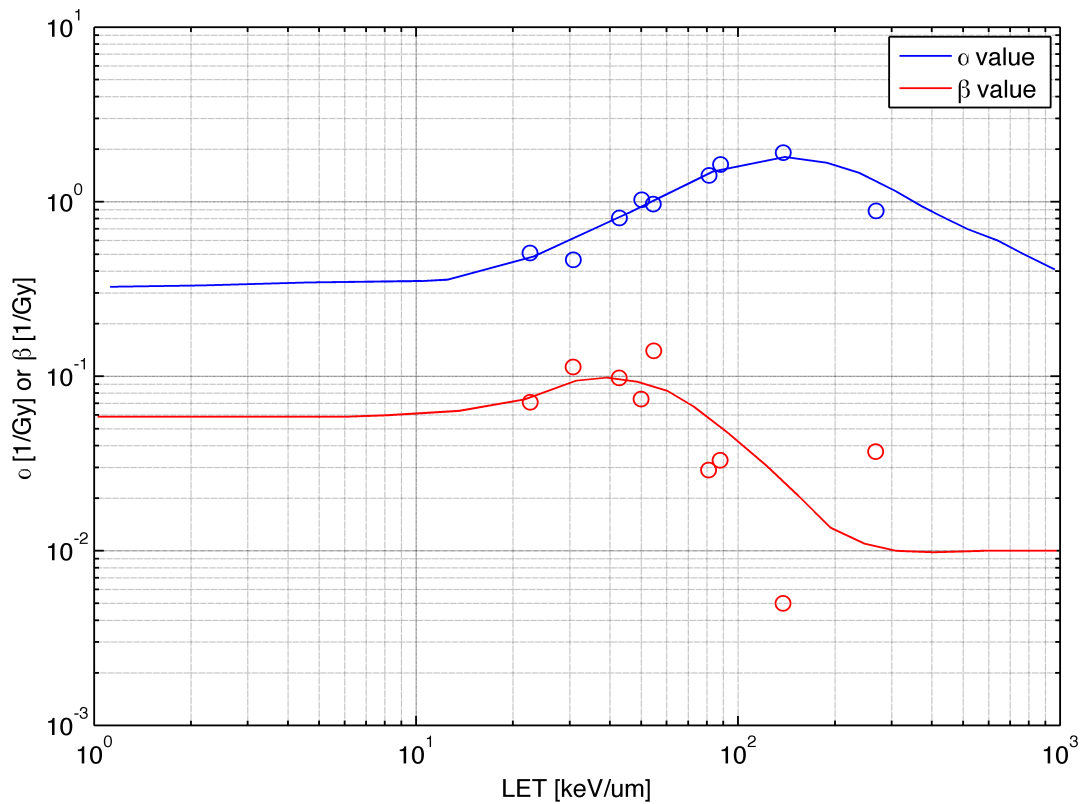

Fig 2. LQ Parameters, alpha and beta as a function of LET

This LQ model was used for RBE calculation. The definition of RBE is well known as a ratio between two absorbed dose delivered with two radiation qualities, one of which is a 'reference radiation'. The gamma ray of $^{60}$Co is used in general [14].

II.3 Spread-out Bragg peak design procedure

In the case of using the Spread-out Bragg peak (SOBP) in the beam delivery, the dose at a position is overlapped by the different depth dose distribution caused by varying incident energy. Therefore, the LET, which was generated by monochromatic carbon energy, cannot use to RBE calculation. Therefore, the concept of dose averaged LET is used [13]. The dose, $D_{SOBP}(x)$, and dose averaged LET, $LET_{SOBP}(x)$, at the position x in SOBP can be calculated by following equation (3) and (4),

$$D_{SOBP}(x) = \sum_j \omega_j d_j(x) \tag{3}$$

$$LET_{SOBP}(x) = \frac{\sum_j LET_j(x)\omega_j d_j(x)}{D_{SOBP}(x)} \tag{4}$$

$d_j(x)$, $w_j(x)$ and $LET_j(x)$ are representing the dose profile, weighting factor and linear energy transfer value from the $j^{th}$ incident beam energy at position $x$. After obtained the dose average LET, it can be used alpha and beta determination on fitted LQ parameters, which was shown in fig 2. Then, the dose for 10% of cell survival fraction can be generated as shown in equation (5).

$$0.1 = \exp(-\alpha D_{10} - \beta D_{10}^2) \tag{5}$$

Then, the RBE based on 10 % of survival fraction can be expressed by equation (6),

$$RBE_{10} = \frac{D_{10 \cdot ref}}{D_{10}} = \frac{4.08 \times 2\beta}{-\alpha + \sqrt{\alpha^2 - 4\beta \ln 0.1}} \tag{6}$$

Then, the biological dose could be determined by multiplication with physical dose and calculated $RBE_{10}$ value. This calculation should be completed through the whole longitudinal distance. In this study, the determination of weighting factor on $j^{th}$ dose profile, $d_j$, would be a key parameter to determine an appropriate SOBP of biological dose. Based on this calculation, the biological SOBP design is followed the work flow diagram, which is shown in Fig 3.

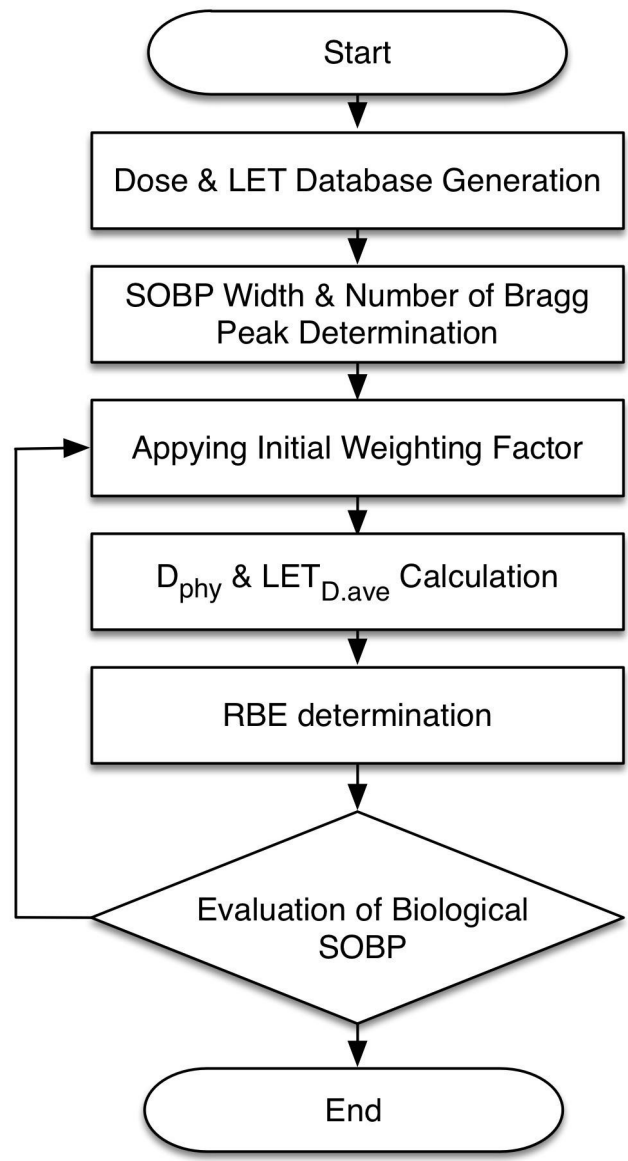

Fig 3. Work flow for design a biological SOBP

II.4 Result

The physical dose and dose averaged LET was calculated from the maximum energy of 340 MeV/u to 250 MeV/u of carbon beam. For the biological SOBP generation, a set of 200 weighting factors was considered, which can be determined the shape of ridge filter [15]. The target SOBP was 10 cm from the maximum Bragg peak position was 20.96 cm.

The weighting factors were applied through several ways to meet the uniformity requirement of biological SOBP as 2.5%. At this study, a set of weighting factors, which is a power function in terms of energy step, were used. The applied weighting factors were shown in Fig 4.

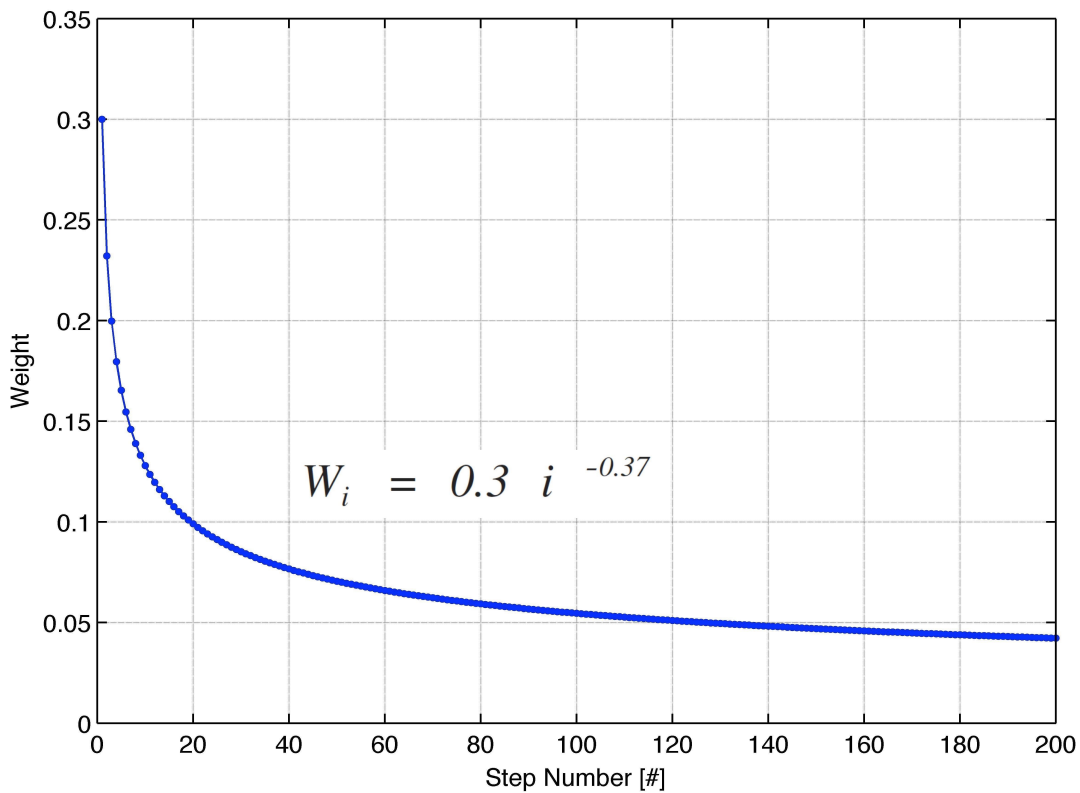

Fig 4. Applied a set of weighting factors

$$W_i = 0.3\ i^{-0.37}$$

As described above, the weighting factor applied to physical dose at each slice, And the LQ parameters were extracted based on the dose-averaged LET, and the RBE for each slice position can be determined. Then, the biological dose profile can be generated by applying RBE to physical dose profile at each position. The biological dose profile uniformity at SOBP region show 1.34%. The weighing factor applied physical dose, which was indicated as blue line, and RBE applied biological dose, which was drawn as red line, were shown in Fig 5.

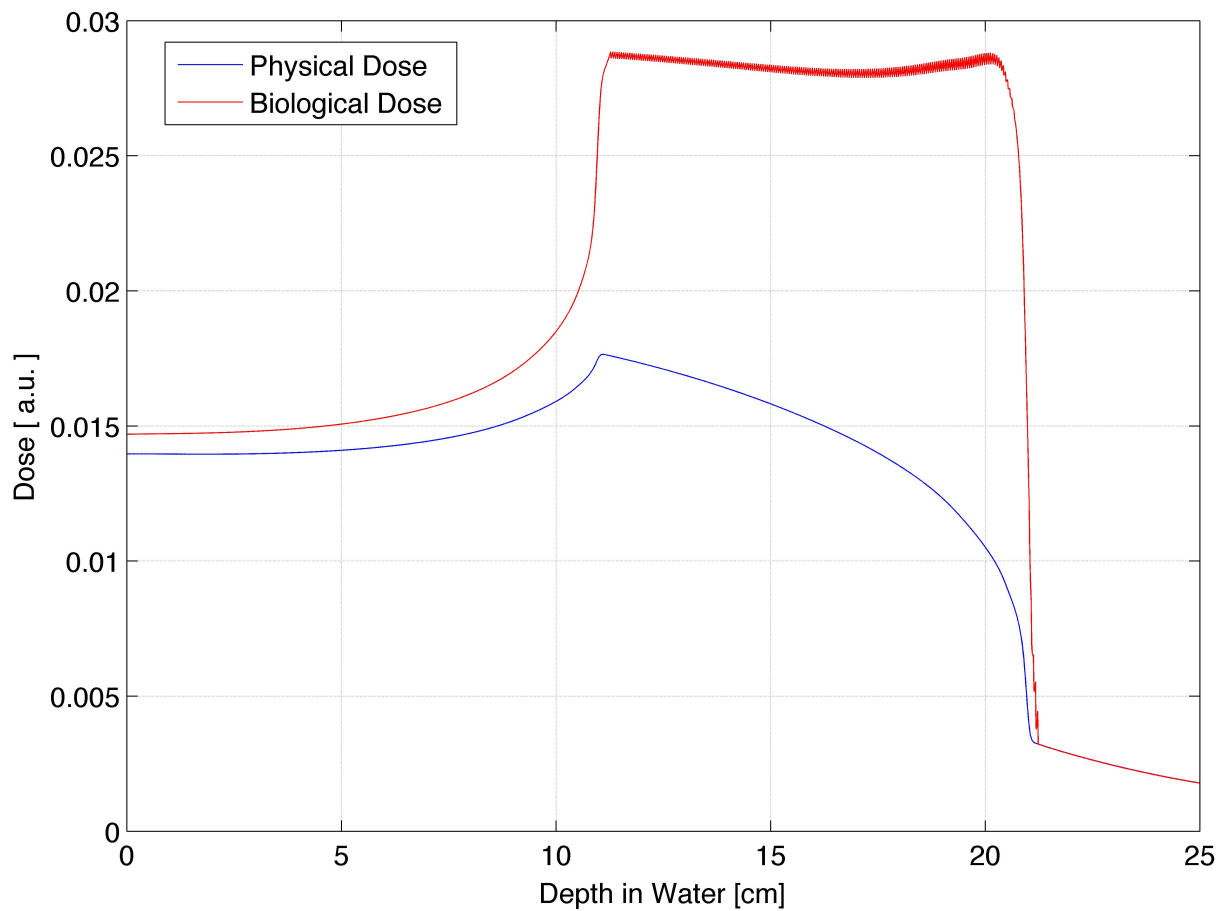

Fig 5. Obtained biological SOBP and weighting applied physical dose

## III. Conclusion

The carbon beam is classified high LET radiation and it causes higher cell killing effects compared other proton and photon. Therefore, the concept of RBE is one of important issue to determine dose on the target. Also, in case of using SOBP, the dose profiles were overlapped each other with different weighting factors. Therefore, the process of RBE determination on a position x became more complex. At this study, the biological SOBP was generated by using LQ model and simulation of carbon beam. The simulation study was focused on obtaining the response of carbon beam at the water phantom. The maximum energy of carbon beam was set to 340 MeV/u on the simulation. The depth dose profile and LET were obtained on the longitudinal direction with 200 different energies, which is corresponding 10 mm of SOBP width, by modifying the Geant4 hadrontherapy example code. To determine RBE on each phantom slice position, the LQ model, based on NIRS experiments data, was applied. A set of weighting factors, which is a power function of applied energy step, could have achievable an appropriate biological SOBP in simple way. The uniformity of the biological SOBP was shown as 1.34%, which met the uniformity requirement [16].

## Acknowledgement

This work was supported by the National Research Foundation of Korea (NRF) grant funded by the Korea government's Ministry of Science, ICT and Future Planning (MSIP)(2015001637)

## Reference

[1] R.R. Wilson, Radiology, **47**, 487 (1946).

[2] D.W. Miller, Med. Phys., **22**, 1943 (1995).

[3] http://ptcog.web.psi.ch/Archive/pat-statistics/Patients tatistics-updateMar2013.pdf.

[4] D.Schulz-Ertner and H.Tsujii, Journal of Clinical Oncology, **25**, 953 (2007).

[5] G. Kraft, Strahlenther Onkol, **175**, 44 (1999).

[6] T.E. Schultheiss, G.K. Zagars, and L.J. Peters, Radiotherapy and Oncology, **9**, 241 (1987).

[7] M. Krämer and M. Scholz, Phys. Med. Biol., **45**, 3319 (2000).

[8] A. Uzawa, K.Ando, S.Koike, Y.Furusawa, Y.Matsumoto, N.Takai, R.Hirayama,

M.Watanabe, M.Scholz, T.Elsässer, P.Peschke, Int. J. Radiation Oncology Biol. Phys., **73**, 1545 (2009).

[9] N. Matsufuji, *Carbon-Ion Radiotherapy* (Springer Japan, Tokyo, 2014), Ch 5, pp. 39–45.

[10] http://www.lns.infn.it/link/Hadrontherapy.

[11] G. Cirrone, G. Cuttone, and S.E. Mazzaglia, Prog. Nucl. Sci., **2**, 207 (2011).

[12] D.J. Brenner, Seminars in Radiation Oncology, **18**, 234 (2008).

[13] Y. Kase, N. Kanematsu, T. Kanai, and N. Matsufuji, Phys. Med. Biol., **51**, N467 (2006).

[14] K.Kagawa, M.Murakami, Y.Hishikawa, M.Abe, T.Akagi, T.Yanou, G.Kagiya, Y.Furusawa, K.Ando, K.Nojima, M.Aoki, T.Kanai, Int.J. Radiat. Oncol. Biol. Phys., **54**, 928 (2002)

[15] C.H. Kim, G. Han, H.-R. Lee, H. Kim, H.S. Jang, J.-H. Kim, D.-W. Park, S.D. Jang, W.T. Hwang, G.-B. Kim, and T.-K. Yang, J.Korean Phys. Soc., **64**, 1308 (2014).

[16] M. Torikoshi, S. Minohara, and N. Kanematsu, J. Radiat. Res., **48**, A15 (2007).